%% file: Circular-Main.tex
\documentclass[a4paper,fleqn]{cas-dc}
\usepackage[numbers]{natbib}
\usepackage{threeparttablex, booktabs, url, amsmath, xr, subfiles}

\def\tsc#1{\csdef{#1}{\textsc{\lowercase{#1}}\xspace}}
\tsc{WGM}
\tsc{QE}

\begin{document}

\shorttitle{Feng et al.}
\title [mode = title]{Circular Polarization in two Active Repeating Fast Radio Bursts}  



%

\author[1]{Yi Feng}

\fnmark[1]





\author[2,1,3]{Yong-Kun Zhang}
\fnmark[1]

\author[2,1,3,4]{Di Li}
\cormark[1]
\ead{dili@nao.cas.cn}

\author[5,6]{Yuan-Pei Yang}

\author[2,7]{Pei Wang}

\author[8,2]{Chen-Hui Niu}

\author[9]{Shi Dai}

\author[10]{Ju-Mei Yao}





\affiliation[1]{organization={Research Center for Intelligent Computing Platforms},
            addressline={Zhejiang Laboratory}, 
            city={Hangzhou},
            citysep={}, 
            postcode={311100}, 
            country={China}}

\affiliation[2]{organization={National Astronomical Observatories},
            addressline={Chinese Academy of Sciences}, 
            city={Beijing},
            citysep={}, 
            postcode={100101}, 
            country={China}}
            
\affiliation[3]{organization={University of Chinese Academy of Sciences},
            city={Beijing},
            citysep={}, 
            postcode={100049}, 
            country={China}}

\affiliation[4]{organization={NAOC-UKZN Computational Astrophysics Centre},
            addressline={University of KwaZulu-Natal}, 
            city={Durban},
            citysep={}, 
            postcode={4000}, 
            country={South Africa}}
            
\affiliation[5]{organization={South-Western Institute for Astronomy Research},
            addressline={Yunnan University}, 
            city={Kunming},
            citysep={}, 
            postcode={650500}, 
            state={Yunnan},
            country={China}}

\affiliation[6]{organization={Purple Mountain Observatory},
            addressline={Chinese Academy of Sciences}, 
            city={Nanjing},
            citysep={}, 
            postcode={210023}, 
            state={Jiangsu},
            country={China}}   
            
\affiliation[7]{organization={Institute for Frontiers in Astronomy and Astrophysics},
            addressline={Beijing Normal University}, 
            city={Beijing},
            citysep={}, 
            postcode={102206}, 
            country={China}}

\affiliation[8]{organization={Institute of Astrophysics},
            addressline={Central China Normal University}, 
            city={Wuhan},
            citysep={}, 
            postcode={430079}, 
            state={Hubei},
            country={China}}
            
\affiliation[9]{organization={School of Science, Western Sydney University},
            addressline={Locked Bag 1797}, 
            city={Penrith NSW},
            citysep={}, 
            postcode={2751}, 
            country={Australia}}
            
\affiliation[10]{organization={Xinjiang Astronomical Observatory},
            addressline={Chinese Academy of Sciences}, 
            city={Urumqi},
            citysep={}, 
            postcode={830011}, 
            country={China}}

\cortext[1]{Corresponding author}
\fntext[fn1]{These authors contributed equally: Y. Feng, Y.K. Zhang.}

\maketitle




Fast radio bursts (FRBs) are bright millisecond-duration radio transients first discovered by \cite{2007Sci...318..777L}. While their cosmological origin and energetic nature make them ideal tools for probing a range of astrophysics \cite{2020Natur.581..391M}, their progenitors and radiation mechanisms are still unknown. A particularly interesting subset of FRBs is the so-called repeating FRBs, which recurrently emit millisecond-duration radio bursts.

Polarization is a fundamental property of FRBs. Faraday rotation measure (RM) carries critical information about the intervening and circumburst environments. The polarization angle and degree of linear and circular polarization can be used to trace the radiation mechanisms and propagation processes \cite{2019MNRAS.483..359L}. For example, the polarization angle of FRB~20180301A showed various short-time-scale swings, which is hypothesized to originate within the magnetosphere of a magnetar \cite{luo2020}. Circular polarization has been detected in about half of non-repeating FRBs \cite{cho20, askap20, feng22}\footnote{See Table S1 in \cite{feng22} for a summary of available circular polarization measurements of non-repeating FRBs.}, for which the polarization was detectable. Linear polarization has been detected in almost all repeating FRBs. In contrast, circular polarization is only seen in one repeating source FRB~20201124A \cite{2021MNRAS.508.5354H}. 

Out of more than 600 published FRBs, only two, namely FRB 20121102A and FRB 20190520B, are found to coincide with a compact persistent radio source (PRS).  We also revealed  their extreme activity \cite{li21,niu22} and significant frequency evolution of linear polarization \cite{feng22}. These facts suggest that these two sources are in complex plasma environment, either young in FRBs' evolution or a special sub-population of FRBs.  

In this study, we reported new detections of circular polarization of both FRBs~20121102A and 20190520B by the Five-hundred-meter Aperture Spherical radio Telescope (FAST) \cite{2011IJMPD..20..989N}, thus tripling the size of repeating FRB sample with circular polarization.

\begin{figure*}[!htp]
 \centering
 \includegraphics[width=0.8\textwidth]{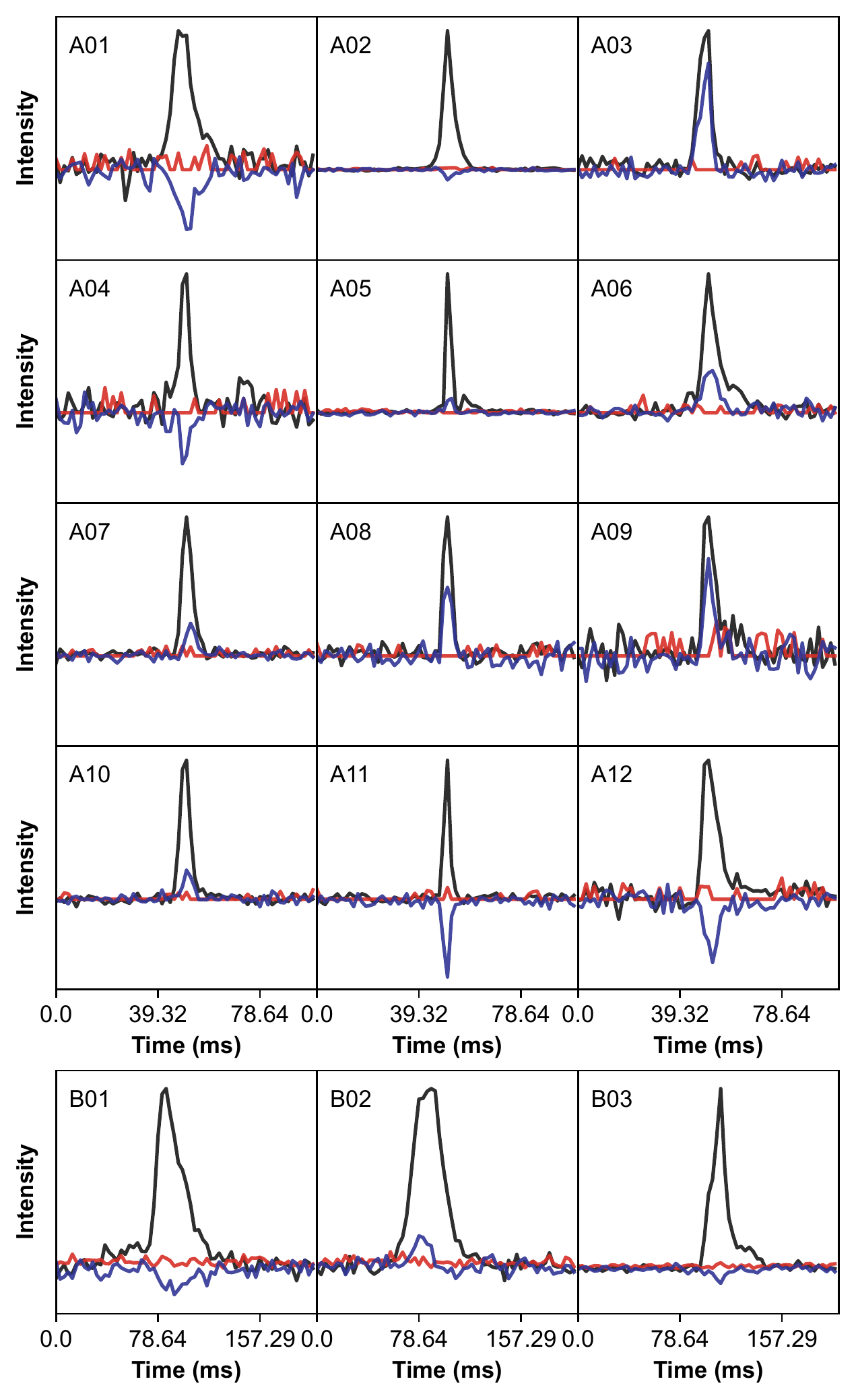}
 \caption{Polarization profiles of twelve bursts with circular polarization of FRB~20121102A (A 01-12) and three bursts of FRB~20190520B (B 01-03) with FAST. The black, red and blue lines represent Stokes I, the linear polarization and Stokes V, respectively.}
 \label{fig:121102}
\end{figure*}

FRB~20121102A is the first precisely-localized repeating FRB \cite{2017Natur.541...58C}. FRB~20121102A has almost 100\% linear polarization at 4-8\,GHz \cite{121102rm}. In contrast, it has no linear polarization at 1.25\,GHz \cite{li21, feng22} thus no measurable RM. The depolarization toward lower frequencies can be well explained by RM scatter due to  multipath propagation \cite{feng22,Yang2022}. In fact, such frequency evolution of polarization seems to be a unified feature of all active repeaters \cite{feng22}. FAST detected 1652 independent bursts  in 59.5 hours spanning 62 days \cite{li21}, resulting in a bimodal energy distribution. Further analyses of the same dataset revealed circular polarization in twelve bursts. The largest degree of circular polarization is about 64\%. FRB~20190520B is the first persistently active FRB, discovered through the Commensal Radio Astronomy FAST Survey (CRAFTS) \cite{li18} and  then localized by the Karl G. Jansky Very Large Array (VLA)-realfast system \cite{niu22}. Similar to FRB~20121102A, FRB~20190520B has no linear polarization at 1.25\,GHz \cite{feng22} because of RM scatter. Further analyses of the FAST sample in \cite{niu22} revealed circular polarization in three bursts. The details of the observations and data reduction can be found in Supplementary materials A. The time of arrival and the degree of circular polarization of each pulse can be found in Table~S1. The polarization pulse profiles are shown in Figure~\ref{fig:121102}. 

We then present spectral and temporal properties of the circular polarization. We show the dynamic spectra of Stokes V and flux density of Stokes V over frequency for six bursts with significant circular polarization in Figure~S1. We did not detect any oscillation or sign change of circular polarization over frequency. In our sample, the circular polarization remained rather constant during the duration of any single burst, no sign change nor other significant variation. For example, we show degrees of circular polarization across $\sim$2\,ms of burst 3, 8, 11 of FRB~20121102A in Figure~S2. The degrees of circular polarization remain relatively constant in $\sim$\,ms time-scale and the variations are within the ranges of error bars.

We consider two categories of mechanisms for generating circular polarization, namely processes during propagation versus radiation mechanism intrinsic to the FRB source. During propagation, multipath propagation and Faraday conversion \cite{2019MNRAS.485L..78V, 2019ApJ...876...74G} could, in some circumstances, generate circular polarization. Multipath propagation occurs when the electromagnetic radiation propagates in an inhomogeneous magneto-ionic environment. As these two FRBs have the largest RM scatter \cite{feng22} corresponding to substantial surrounding electron density and complexities, they would have a better chance to have multipath propagation induced circular polarization than other FRBs. However, we demonstrated that the observed significant frequency-averaged circular polarization is unlikely induced by multipath propagation (Supplementary materials B). Faraday conversion is a relatively weak effect and generates observable circular polarization only when propagating through an extremely magnetized region with magnetic field reversals or propagating through strongly magnetized plasma consisting of  relativistic electrons. As the two FRBs have complex environments, Faraday conversion could take place, but should remain rare because only a small fraction of bursts have observable circular polarization.
 
Finally, we consider radiation mechanism intrinsic to the FRB source. The observed circular polarization may originate within the magnetosphere of a magnetar, an increasingly favored origin of FRBs.  Circular polarization is commonly seen in pulses from magnetars. The rarity of   circularly polarized FRB bursts indicate similar conditions should be rare in FRB sources, even if the generation of circular polarization goes through analogous processes. We note that there are some variations on circular polarization between two adjacent bursts, as shown in Table 1. The difference of circular polarization might be due to the different magnetic field configurations or different beaming directions deviating from the line of sight, if the radiation mechanism is the coherent curvature radiation \cite{2022ApJ...927..105W, 2022RAA....22g5013T, 2022MNRAS.517.5080W}. 
 
We then compare the circular polarization properties of these two active repeating FRBs with those of non-repeating FRBs. It seems that the circular polarization of the non-repeating FRBs is different from the repeating FRBs. The variation time-scale of the circular polarization of the non-repeating FRBs can be smaller than 1\,ms. For example, the degree of circular polarization of FRB~20190611B varies from 15\% to 57\% in $\sim$1\,ms \citep{askap20}. The degree of circular polarization of FRB~20181112A varies from -34\% to 17\% in less than $\sim$0.1\,ms \citep{cho20}. The variation time-scale of the circular polarization of the non-repeating FRBs is much smaller than that of the repeating FRBs. The circular polarization of the non-repeating FRBs has been attributed to intrinsic radiation mechanism  or the result of propagation through a relativistic plasma close to the source \citep{cho20}. The short, millisecond-scale variation of the circular polarization of the non-repeating FRBs seems to favor intrinsic processes as they are closer to the presumed compact object. 

Our observations have tripled the size of repeating FRB sample with circular polarization. The observed circular polarization is unlikely induced by multipath propagation. Our observations favor circular polarization induced by Faraday conversion or radiation mechanism intrinsic to the FRB source. The conditions to generate circular polarization have to be rare in either case, as there are only about 1\% and 4\% of bursts being seen with circular polarization for FRB~20121102A and FRB~20190520B respectively, much less than the $\sim$50\% for non-repeating FRBs. Further, systematic study of circular polarization will shed critical light on the environment and radiation mechanisms of repeating FRBs with this growing sample.       
\section*{Competing interests}
The authors declare no competing interests.

\section*{Acknowledgments}
We would like to thank Don Melrose for valuable discussions. 
This work is supported by National Natural Science Foundation of China grant No.\ 11988101, 12203045, 11725313, by Key Research Project of Zhejiang Laboratory no.\ 2021PE0AC03, and by National Key R$\&$D Program of China No. 2017YFA0402600. Yuan-Pei Yang is supported by National Natural Science Foundation of China grant No. 12003028, the China Manned Spaced Project (CMS-CSST-2021-B11), and the National Key Research and Development Program of China (2022SKA0130101). Shi Dai is the recipient of an Australian Research Council Discovery Early Career Award (DE210101738) funded by the Australian Government. Pei Wang acknowledges support from the National Natural Science Foundation of China under grant U2031117, the Youth Innovation Promotion Association CAS (id. 2021055), CAS Project for Young Scientists in Basic Reasearch (grant YSBR-006) and the Cultivation Project for FAST Scientific Payoff and Research Achievement of CAMS-CAS. This work made use of the data from FAST (Five-hundred-meter Aperture Spherical radio Telescope), a Chinese national mega-science facility, operated by National Astronomical Observatories, Chinese Academy of Sciences. 

\section*{Author contributions}
Yi Feng led the data analysis, interpretations and manuscript preparation. Yong-Kun Zhang led the data visualization. Di Li  launched the FAST observation campaign and contributed to the writing. Pei Wang and  Chen-Hui Niu carried out the data acquisition. Yuan-Pei Yang, Shi Dai and Ju-Mei Yao contributed to theoretical investigations. All authors discussed interpretations and commented on the manuscript.
\printcredits

\section*{Data availability}
The data of the fifteen bursts and the calibration files are openly available in Science Data Bank at \url{https://doi.org/10.57760/sciencedb.04389}.

%
\bibliographystyle{model3-num-names}
\bibliography{cas-refs}

\clearpage

\bio{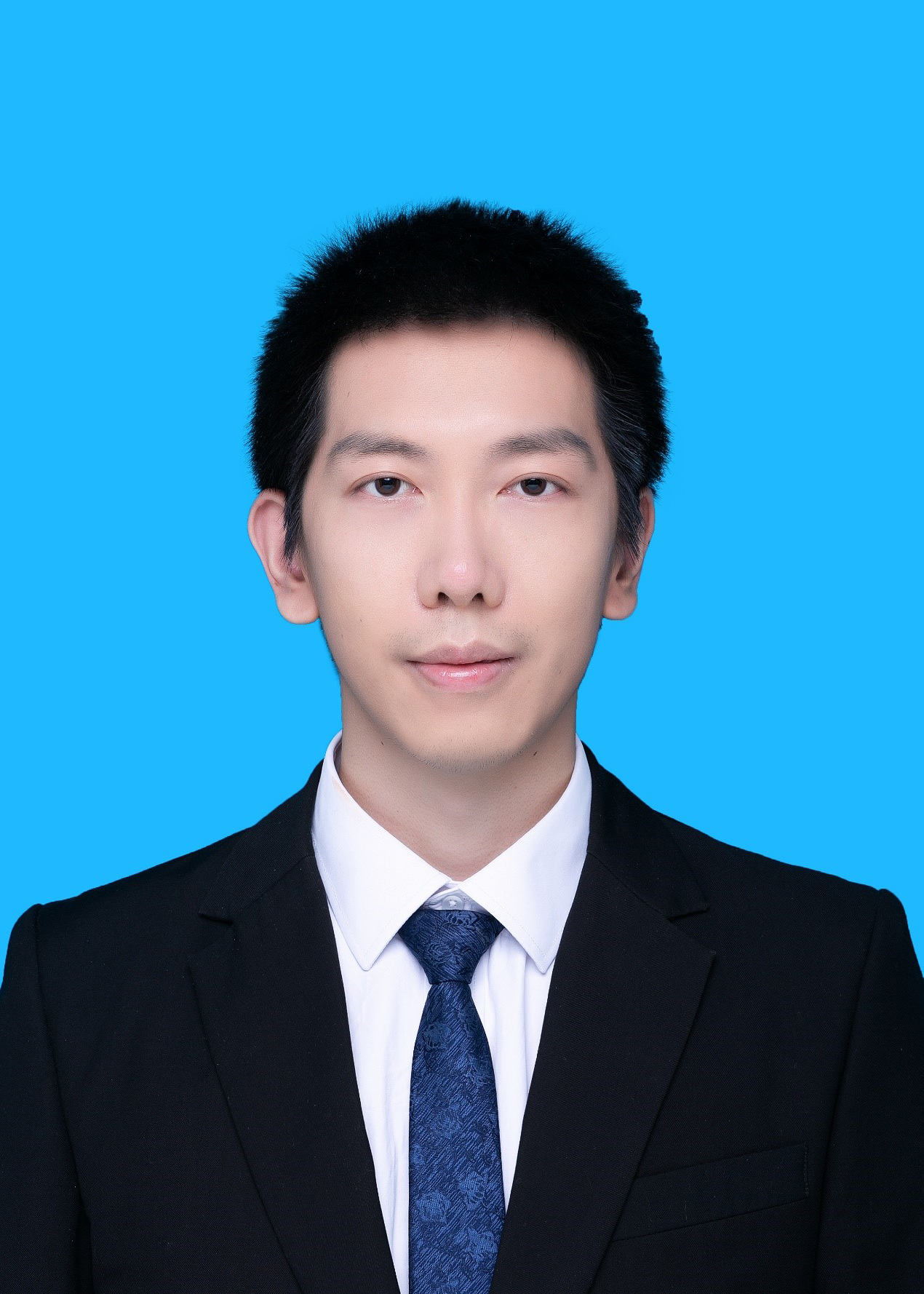}
Yi Feng is a research fellow at the Zhejiang Laboratory. He received his Bachelor’s degree in Physics from Tsinghua University in 2013 and Ph.D. degree in Astrophysics from University of Chinese Academy of Sciences in 2021. His research focuses on the fast radio bursts, pulsars, and gravitational waves. He has published more than 30 papers, including 5 papers in Nature and Science (two first-author Nature/Science papers).
\endbio

\bio{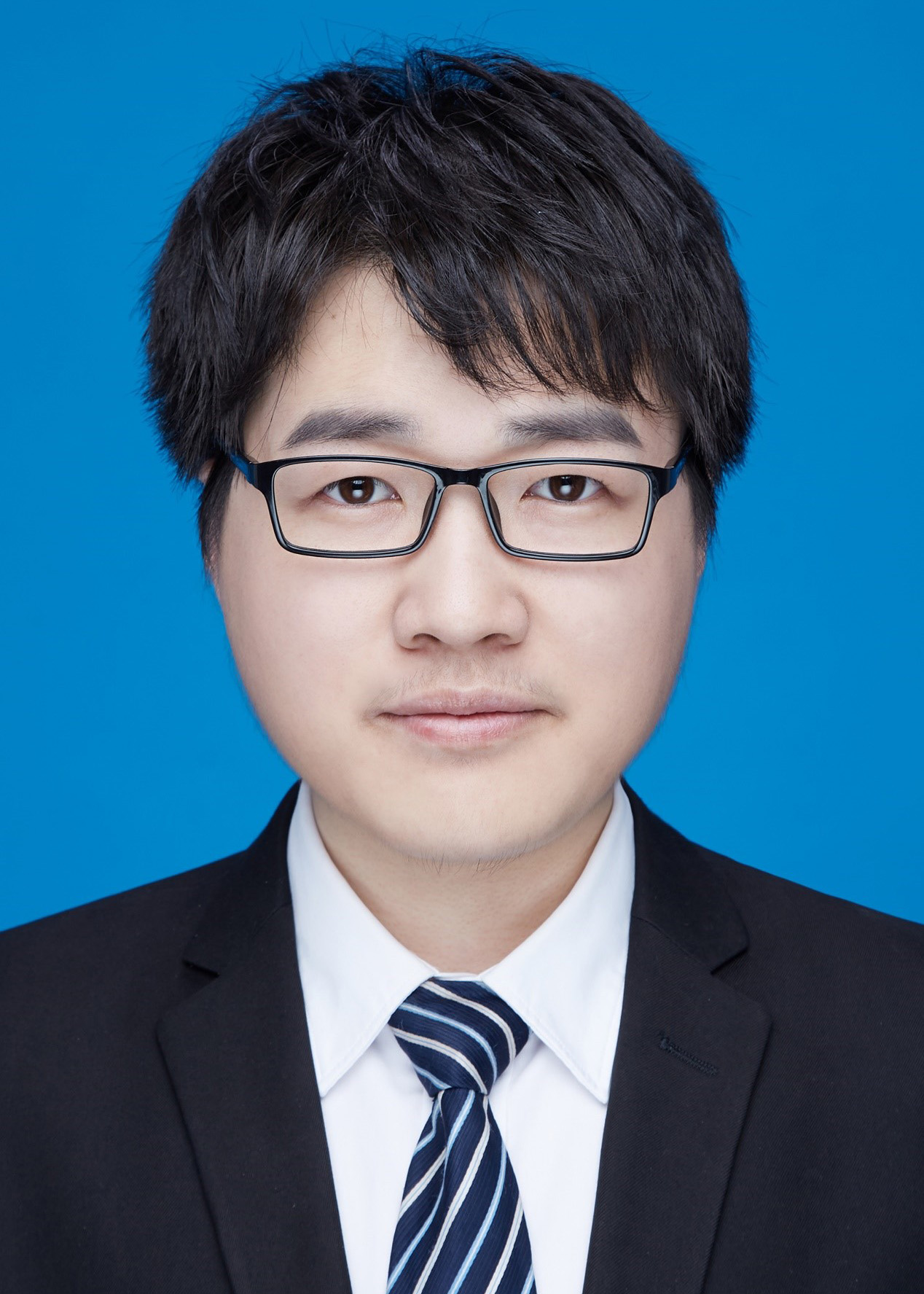}
Yong-Kun Zhang is a Ph.D. candidate at National Astronomical Observatories, Chinese Academy of Sciences. He obtained his Bachelor’s degree in Physics from University of Chinese Academy of Sciences in 2019. His current research interests include fast radio bursts, star formation, and applications of machine learning in astronomy.
\endbio

\vspace{6ex}

\bio{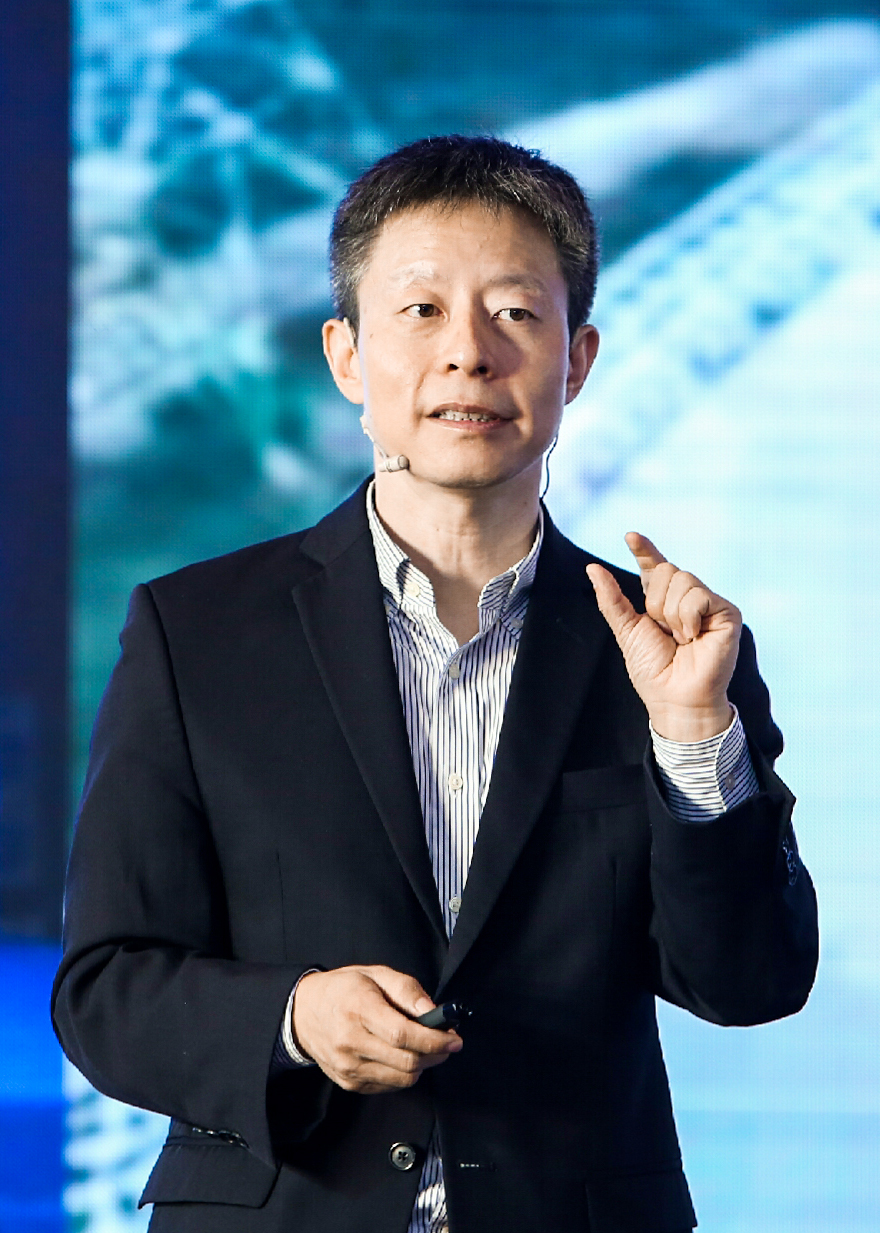}
Di Li is the Chief Scientist of the radio division of the National Astronomical Observatories, Chinese Academy of Sciences. He has been the chief scientist of FAST since 2018. He received his Bachelor’s degree in Physics from Peking University in 1995 and Ph.D. degree in Astrophysics from Cornell University in 2002. His research work includes star formation, fast radio bursts, pulsars, gravitational waves, astrochemistry, and radio astronomy techniques.
\endbio

\clearpage
\subfile{Circular-Supp.tex}

\end{document}

%% file: Circular-Supp.tex
\shorttitle{Feng et al.}
\centerline{\bf\huge Supplementary Text}

\setcounter{section}{0}
\renewcommand{\thesection}{S\arabic{section}}
\setcounter{equation}{0}
\renewcommand{\theequation}{S\arabic{equation}}
\setcounter{figure}{0}
\renewcommand{\thefigure}{S\arabic{figure}}
\setcounter{table}{0}
\renewcommand{\thetable}{S\arabic{table}}

\appendix
\section{Observations and data reduction} 
The FAST observations were conducted using the central beam of the 19-beam receiver, which covers frequencies between 1050 and 1450 MHz with two linear polarizations \cite{li18}. The data streams are processed with the Reconfigurable Open Architecture Computing Hardware–version 2 (ROACH2) signal processor. The output data files are recorded as 8 bit-sampled search mode PSRFITS \cite{2004PASA...21..302H} files with 4096 frequency channels.

Polarization calibration was achieved by correcting for the differential gain and phase between the receptors through separate measurements of a noise diode signal injected at an angle of $45^{\circ}$ from the linear receptors with the single-axis model using the PSRCHIVE software package \cite{2004PASA...21..302H}. This calibration scheme does not correct for leakage. At FAST, the leakage term is better than -46 dB within the full width at half maximum region of the central beam as measured during the FAST engineering phase \cite{FAST19Beam}, which corresponds to systematic errors less than 0.5\%. The cross-polarization leakage of the Robert C. Byrd Green Bank Telescope C-Band receiver when recording linear polarization is $< -32$ dB, and over most of the band it is $< -37$ dB, which corresponds to systematic errors of about 1.4\%. 
Systematic error of degree of circular polarization caused by large zenith angle at FAST is less than 1\% \cite{2022arXiv221003609J}, and our results are not affected by large zenith angle.  
To excise radio frequency interference (RFI), we used the PSRCHIVE software package to median filter each burst in the frequency domain and we also mitigated RFI of each burst manually. 

We use the frequency-averaged, de-biased total linear polarization \cite{2001ApJ...553..341E}:
\begin{equation} \label{eq:L_de-bias}
    L_{{\mathrm{de\mbox{-}bias}}} =
    \begin{cases}
      \sigma_I \sqrt{\left(\frac{L_{i}}{\sigma_I}\right)^2 - 1} & \text{if $\frac{L_{i}}{\sigma_I} > 1.57$} \\
      0 & \text{otherwise} ,
    \end{cases}
\end{equation}
where $\sigma_I$ is the Stokes I off-pulse standard deviation and $L_i$ is the measured frequency-averaged linear polarization of time sample $i$. We defined the degree of circular polarization as ($\Sigma_{i} V_i$)/($\Sigma_{i}I_i$), where the summation is over the bursts, $V_i$ is the measured frequency-averaged circular polarization of time sample $i$, and $I_i$ is the measured frequency-averaged Stokes I.
Defining $I = \Sigma_{i}I_i$ and $V = \Sigma_{i}V_i$, uncertainties on the circular polarization fraction are calculated as:
\begin{equation} \label{eq:uncertainty}
    \sigma_{V/I} = \frac{\sqrt{N+N\frac{V^2}{I^2}}}{I}\sigma_{I},
\end{equation}
where $N$ is the number of time samples of the burst.

\section{Multipath propagation induced circular polarization} \label{sec:cir}
Circular polarization can be induced by multipath propagation \cite{Macquart20002, 2022MNRAS.510.4654B}. The interpretation is as follows: the randomly distorted wavefronts form an interference pattern at the location of the observer. The left- and right-hand circularly polarized wavefronts with different refractive indices become spatially separated after propagating through a rotation measure gradient, which behaves like a Faraday wedge. The interference patterns of the two circular polarizations do not overlap in the observer’s plane, which leads to alternate patches in which one circular polarization and then the other dominates, leading to a nonzero $\langle V^2\rangle$.   

Some repeating FRBs are proposed to be related to a SNR or a wind nebula \cite{Yang2017, 2018ApJ...861..150P, 2018ApJ...868L...4M, 121102rm, feng22, Yang2022}. Dense and magnetised environments like SNR or wind nebula are birefringent medium for the left- and right-hand circularly polarized wavefronts and behave like a Faraday wedge. Such inhomogeneous medium can also cause scintillation. The two conditions, scintillation and rotation measure gradient, are naturally met and we expect that circular polarization can be induced when passing through the local dense and magnetised medium for repeating FRBs. Some repeating FRBs were found to exhibit both
scattering and scintillation. The scintillation pattern in frequency is caused by the Milky Way and the scattering tail is associated with the FRB’s
host galaxy. The two scattering screens, one in the Milky Way and the other in the host galaxy, could generate the observed circular polarization in the repeating FRBs. We now discuss circular polarization generated by the two scattering screens respectively.  

\subsection{Multipath propagation scenario: scattering screen in the Milky Way}
We first assume that the observed circular polarization is induced by the scattering screen in the Milky Way.  
One should observe fluctuations
in circular polarization with a root mean square (rms) degree of circular polarization \cite{Macquart20002} as:
\begin{equation}
\label{eq:circular1}
\frac{V_{\mathrm{rms}}}{I} \sim \frac{d_{\mathrm{so}} \lambda^{3} \nabla_{\perp} \mathrm{RM}}{2\pi r_{\mathrm{diff}}},
\end{equation}
where $d_{\mathrm{so}}$ is the distance between the scattering screen and the observer, $\lambda$ is the wavelength, $\nabla_{\perp} \mathrm{RM}$ is the RM gradient in the transverse plane, and $r_{\mathrm{diff}}$ is the diffractive scale, which can be written as 
\begin{equation}
\label{eq:rdiff}
r_{\mathrm{diff}} = \lambda \sqrt{\frac{d_{\mathrm{fs}}d_{\mathrm{so}}}{c \tau_{s} d_{\mathrm{fo}}}},
\end{equation}
where $\tau_{s}$ is the scattering time-scale, $d_{\mathrm{fs}}$ is the distance between the FRB source and the scattering screen, and $d_{\mathrm{fo}}$ is the distance between the FRB source and the observer. As the FRB sources considered here are extragalactic and $\frac{d_{\mathrm{fs}}}{d_{\mathrm{fo}}} \approx 1$, thus $r_{\mathrm{diff}}$ can be approximated as $\lambda \sqrt{\frac{d_{\mathrm{so}}}{c \tau_{s}}}$.

The requirement for large circular polarization is $\frac{d_{\mathrm{so}} \lambda^{3} \nabla_{\perp} \mathrm{RM}}{2\pi r_{\mathrm{diff}}} \gtrsim 1$. In terms of RM gradient, the requirement can be written as $\nabla_{\perp} \mathrm{RM} \gtrsim \frac{2\pi r_{\mathrm{diff}}}{d_{\mathrm{so}} \lambda^{3}}$, and simplified as:
\begin{equation}
\nabla_{\perp} \mathrm{RM} \gtrsim 7.3\times10^{-11} (\frac{f}{1\,\mathrm{GHz}})^2 (\frac{d_{\mathrm{so}}}{1\,\mathrm{kpc}})^{-\frac{1}{2}} (\frac{\tau_{s}}{0.1\,\mathrm{ms}})^{-\frac{1}{2}}\,\mathrm{rad/m^3}.
\end{equation}
The requirement on RM gradient can be translated to:
\begin{equation}
(\frac{n_{\mathrm{e}}}{1\,\mathrm{cm}^{-3}})(\frac{B_{\parallel}}{\mathrm{mG}}) \gtrsim 2.8\times10^{3} (\frac{f}{1\,\mathrm{GHz}})^2 (\frac{d_{\mathrm{so}}}{1\,\mathrm{kpc}})^{-\frac{1}{2}} (\frac{\tau_{s}}{0.1\,\mathrm{ms}})^{-\frac{1}{2}}\frac{l_{\perp}}{l_{\parallel}},
\end{equation}
where $n_{\mathrm{e}}$ is the electron density in electrons per cubic centimetre, $B_{\parallel}$ is the line-of-sight magnetic field strength, $l_{\perp}$ is the characteristic length scale in the transverse plane, and $l_{\parallel}$ is the characteristic length scale in the line of sight.
Such magneto-ionic condition is hard to meet in the scattering screen in the Milky Way. Therefore, we conclude that the observed circular polarization is not caused by the scattering screen in the Milky Way. The observations of scattering and scintillation also suggests that the scattering is not in the Milky Way, because the observed scintillation bandwidth is much larger than that estimated by the scattering for a single scattering screen.

\subsection{Multipath propagation scenario: Scattering screen in the host galaxy}
We then assume that the observed circular polarization is induced by the scattering screen in the host galaxy. The observed degree of frequency-averaged circular polarization should be smaller than $\sim(f_{\mathrm{scin}}/f_{\mathrm{width}})^{1/2}$ \cite{2022MNRAS.510.4654B}, where $f_{\mathrm{scin}}$ is the scintillation bandwidth, and $f_{\mathrm{width}}$ is the bandwidth of the FRB burst. Different with the scattering screen in the Milky Way, the factor $(f_{\mathrm{scin}}/f_{\mathrm{width}})^{1/2}$ will play a key role to limit the observed frequency-averaged circular polarization for the scattering screen in the host galaxy because of the much smaller $f_{\mathrm{scin}}$.     
Take $f_{\mathrm{width}} = 100$\,MHz for typical FRB bandwidth, $\tau_{s} = 0.1$\,ms, $f_{\mathrm{scin}} \sim 1/\tau_{s} \sim 10$\,kHz, and the observed degree of circular polarization should be smaller than 0.01. Thus we conclude that the observed significant frequency-averaged circular polarization is unlikely caused by the scattering screen in the host galaxy. To summarize, the observed circular polarization in the two active repeating FRBs is unlikely induced by multipath propagation. 

\section{Extended data figures and tables}

\begin{table}
\begin{threeparttable}[b]
  \caption{{\bf The degree of circular polarization.}}\label{tab:burst}
  \centering
  \begin{tabular}{ccr}
    \toprule
    Burst & Modified Julian date\tnote{a}  & \%~Circular \\
    \midrule
    \multicolumn{2}{l}{FRB~20121102A} \\
    \hline
    1  & 58725.96433279    & $-33.3\pm3.4$    \\
    2  & 58727.98962489    & $-6.8\pm0.5$     \\
    3  & 58728.00174170    & $61.6\pm2.6$     \\
    4  & 58730.90048725    & $-34.3\pm6.6$    \\
    5  & 58730.90700850    & $14.6\pm1.2$     \\
    6  & 58748.91281354    & $17.5\pm2.0$     \\
    7  & 58748.93111854    & $30.7\pm2.8$     \\
    8  & 58750.86593678    & $52.8\pm5.7$     \\
    9  & 58750.86884520    & $64.0\pm5.9$     \\
    10 & 58755.01306475    & $18.0\pm1.7$     \\
    11 & 58756.84376467    & $-57.4\pm3.4$    \\
    12 & 58759.94263005    & $-35.8\pm3.7$    \\
    \toprule
    \multicolumn{2}{l}{FRB~20190520B} \\
    \hline
    1 & 59061.53656886     & $-12.8\pm1.7$   \\
    2 & 59075.45486247     & $10.7\pm1.2$    \\
    3 & 59077.48545184     & $-7.6\pm0.9$    \\
    \toprule
  \end{tabular}
  \begin{tablenotes}
    \item [a] Arrival time of burst peak at the solar system barycenter, after correcting to the frequency of 1.5\,GHz.
  \end{tablenotes}
\end{threeparttable}
\end{table}


\begin{figure*}[!htp]
  \centering
  \includegraphics[width=0.46\textwidth]{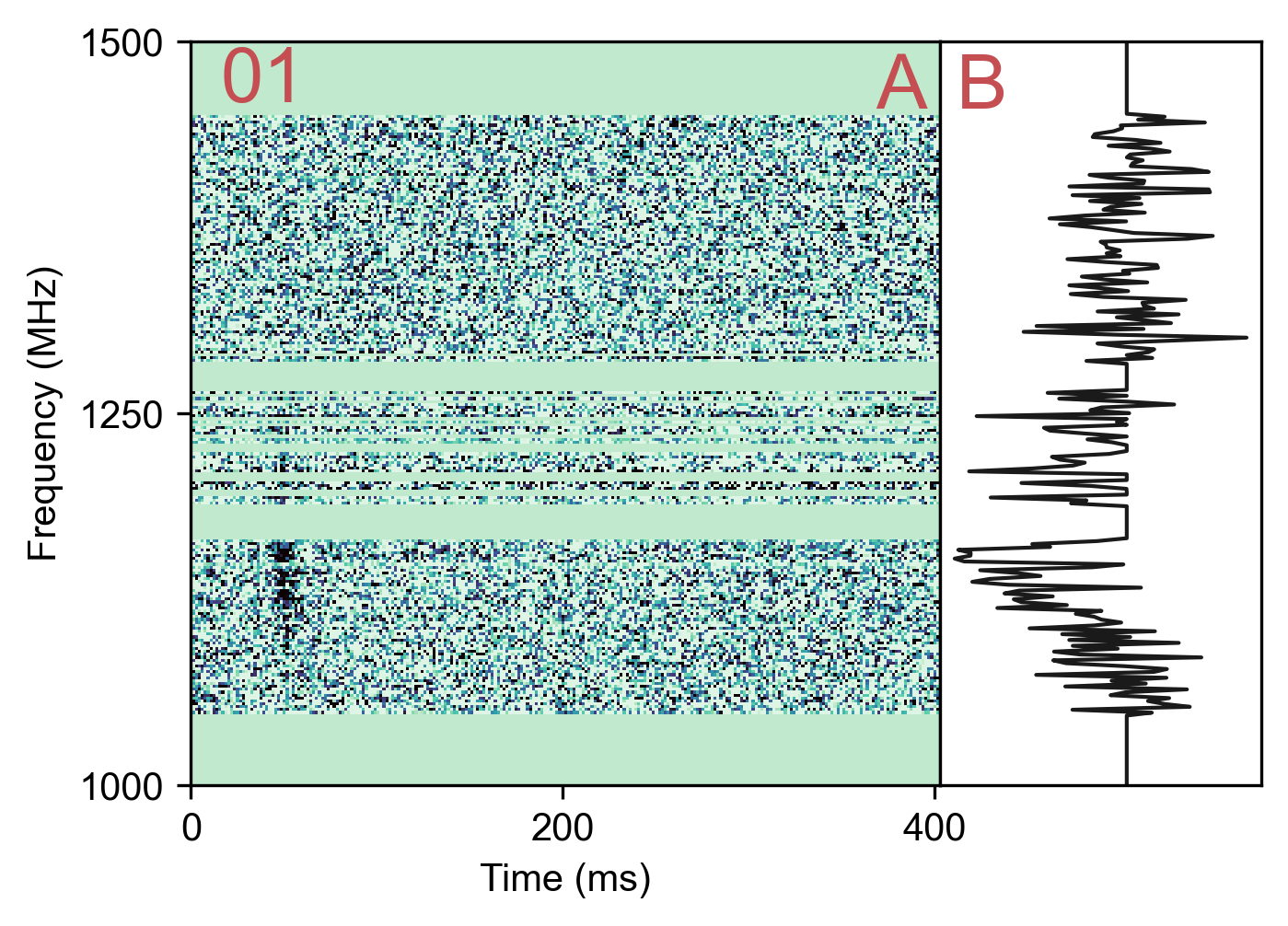}
  \includegraphics[width=0.46\textwidth]{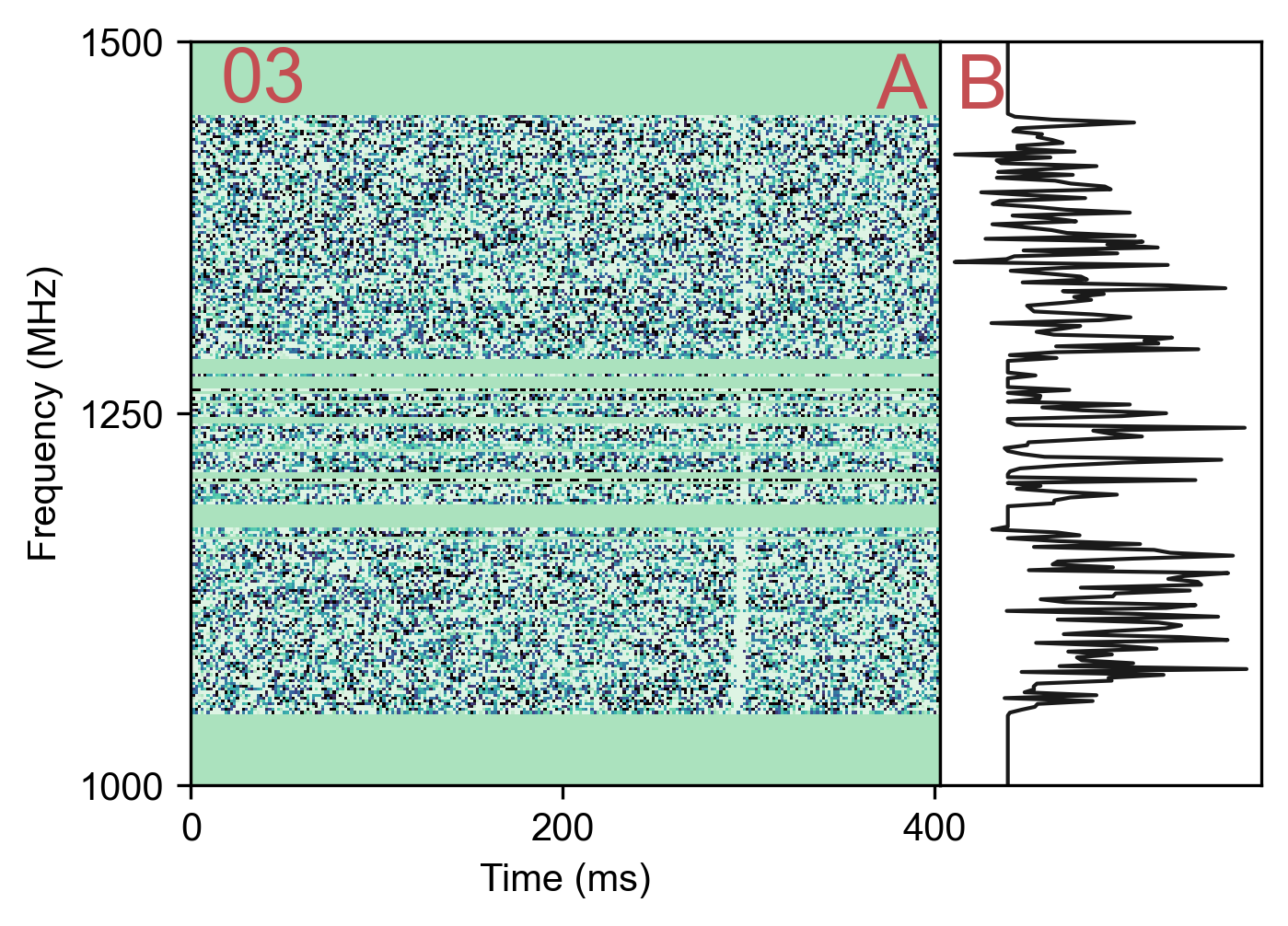}\\
  \includegraphics[width=0.46\textwidth]{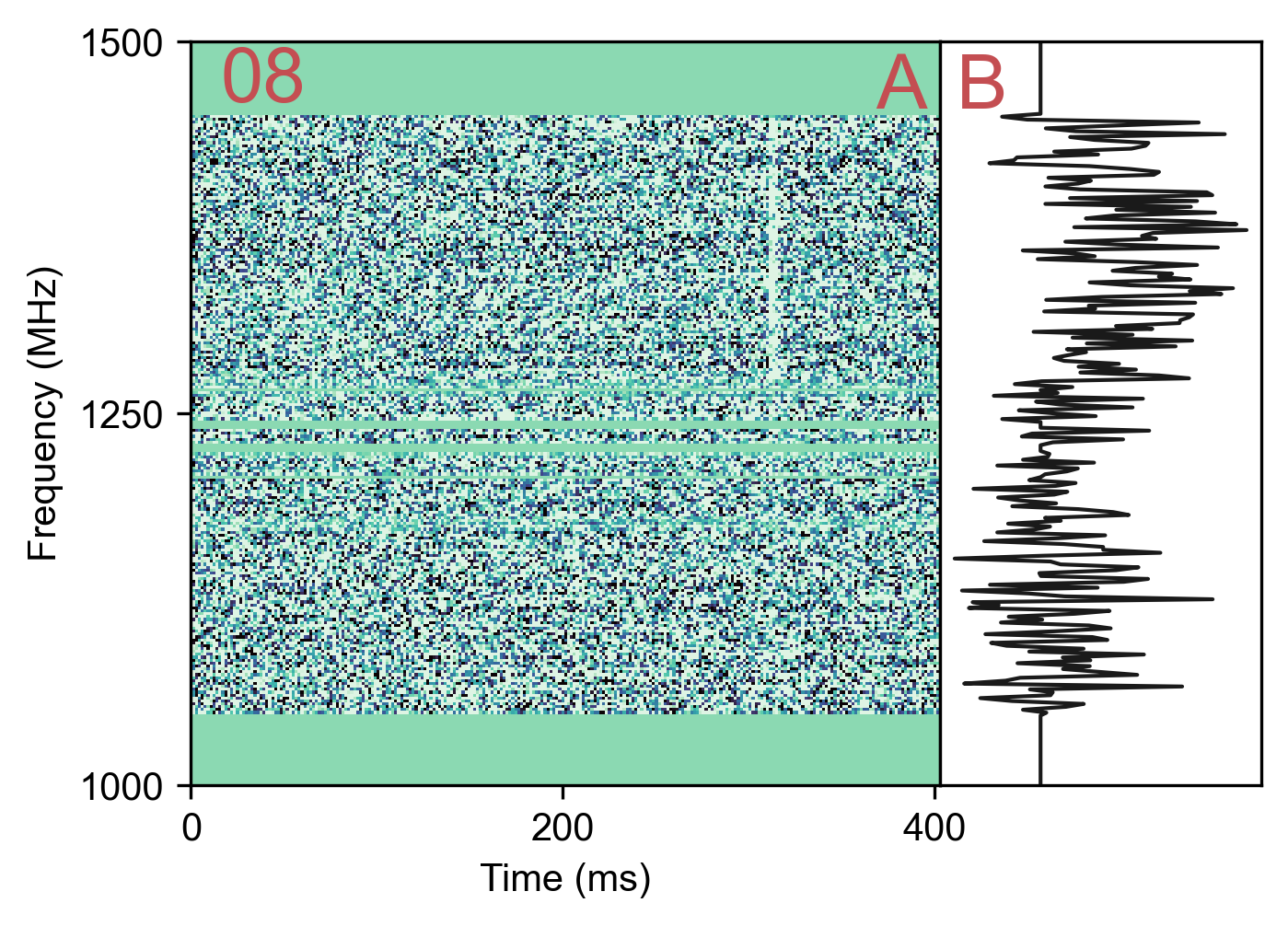}
  \includegraphics[width=0.46\textwidth]{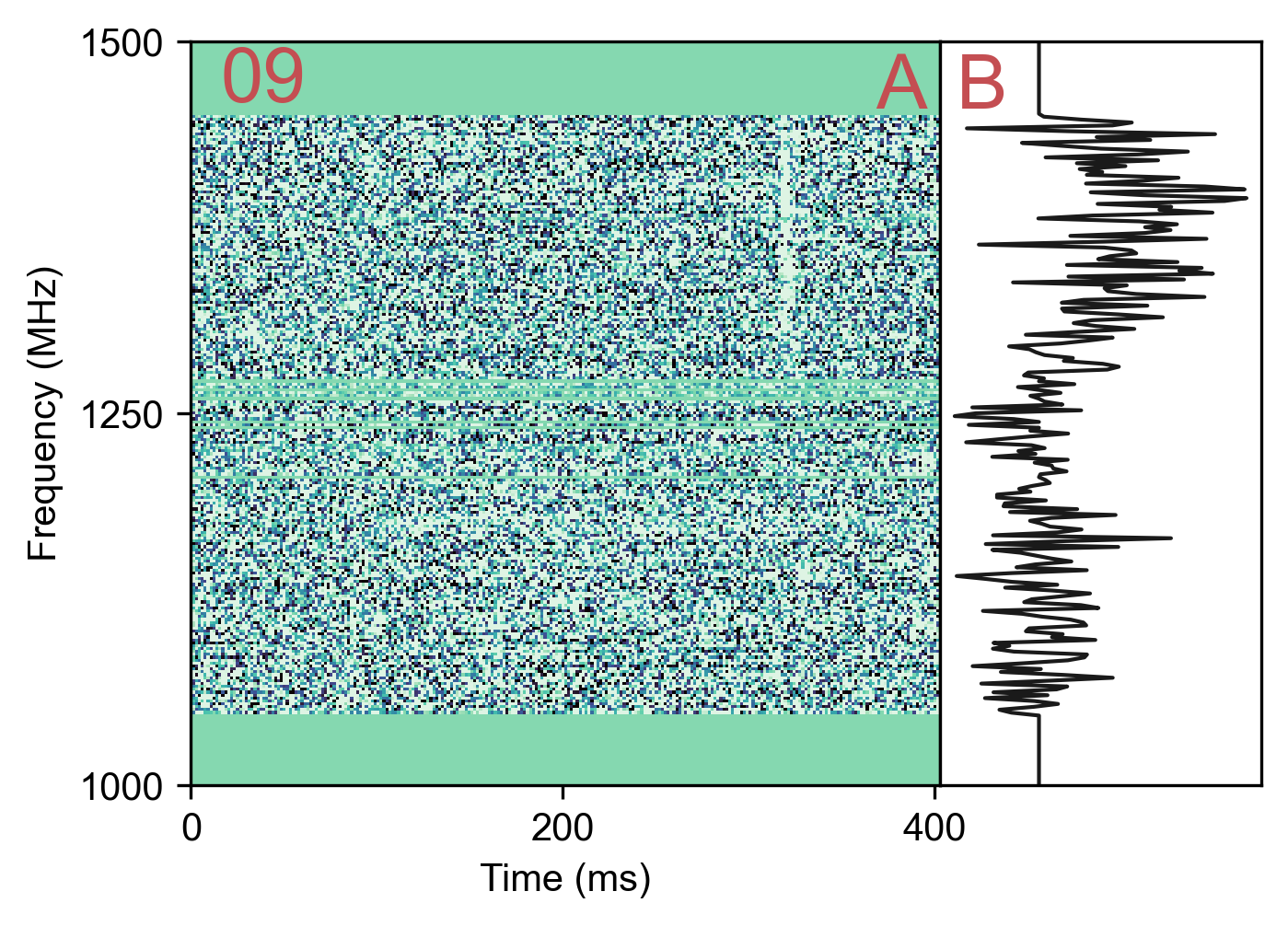}\\
  \includegraphics[width=0.46\textwidth]{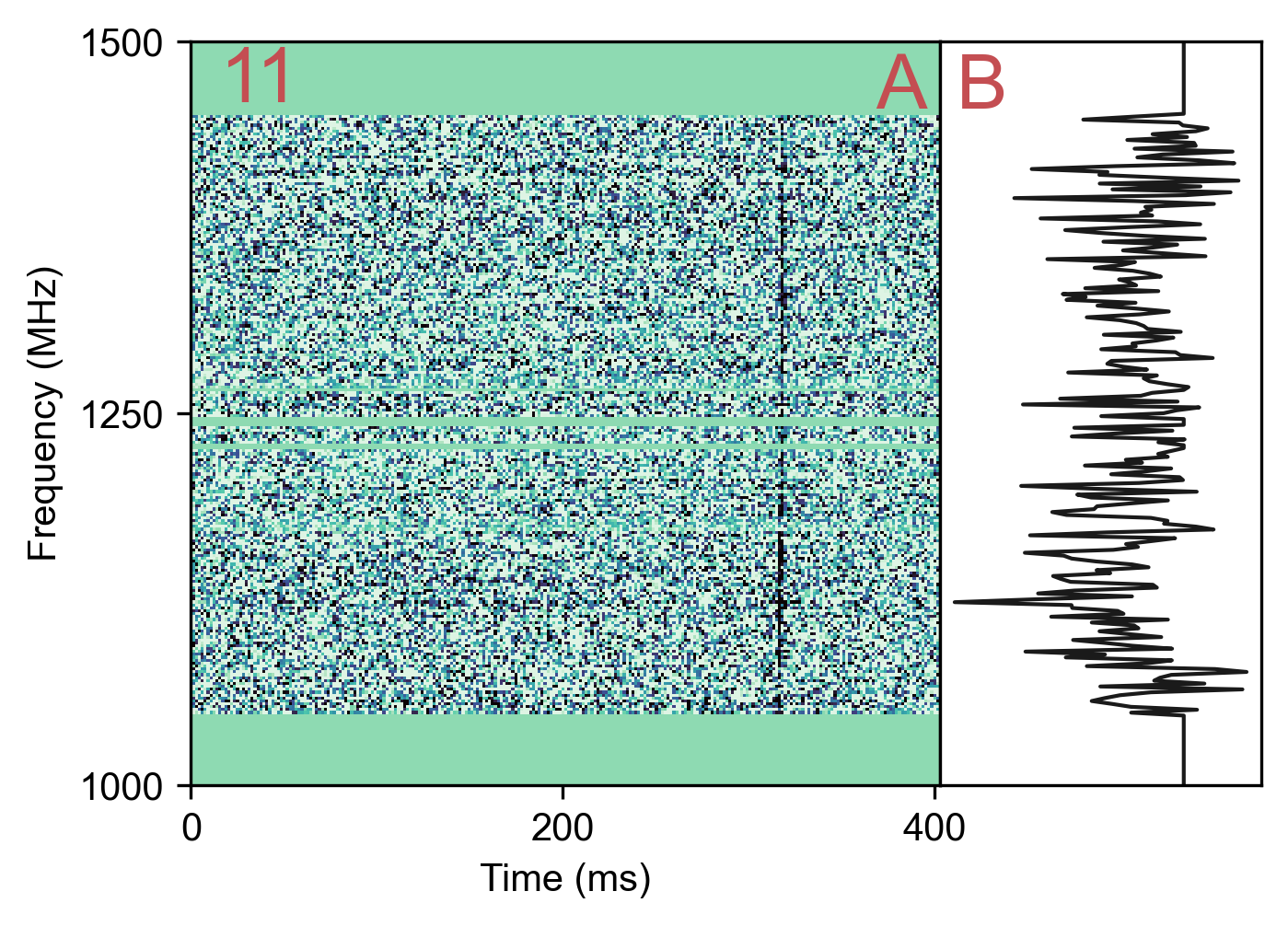}
  \includegraphics[width=0.46\textwidth]{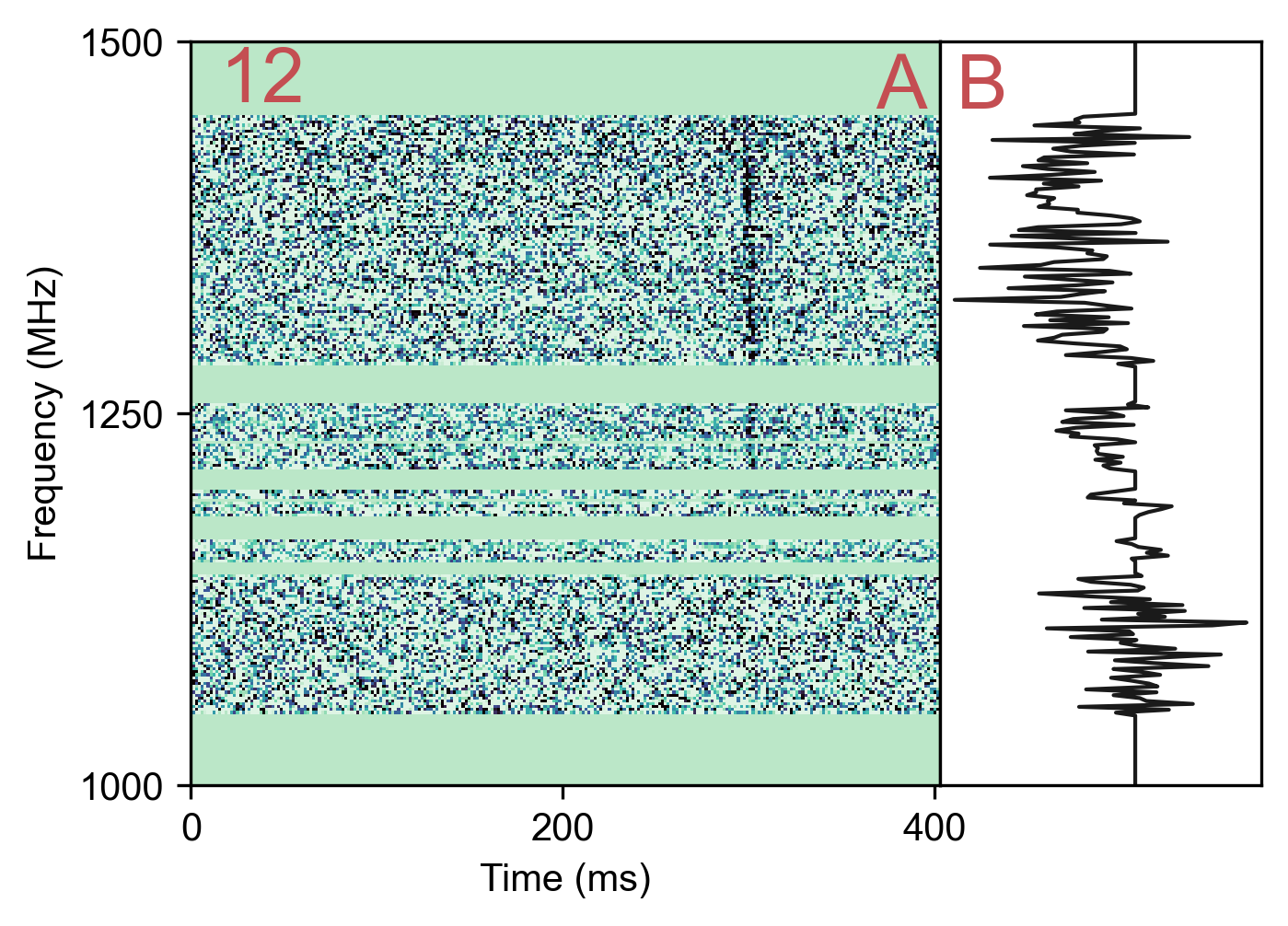}
  \caption{Left sub-panel A: dynamic spectra of Stokes V for six bursts from FRB~20121102A in Figure 1 with burst index shown in the top left. Right sub-panel B: corresponding flux density of Stokes V over frequency in sub-panel A.}\label{fig:spec}
\end{figure*}

\begin{figure}
  \centering
  \includegraphics[width=0.46\textwidth]{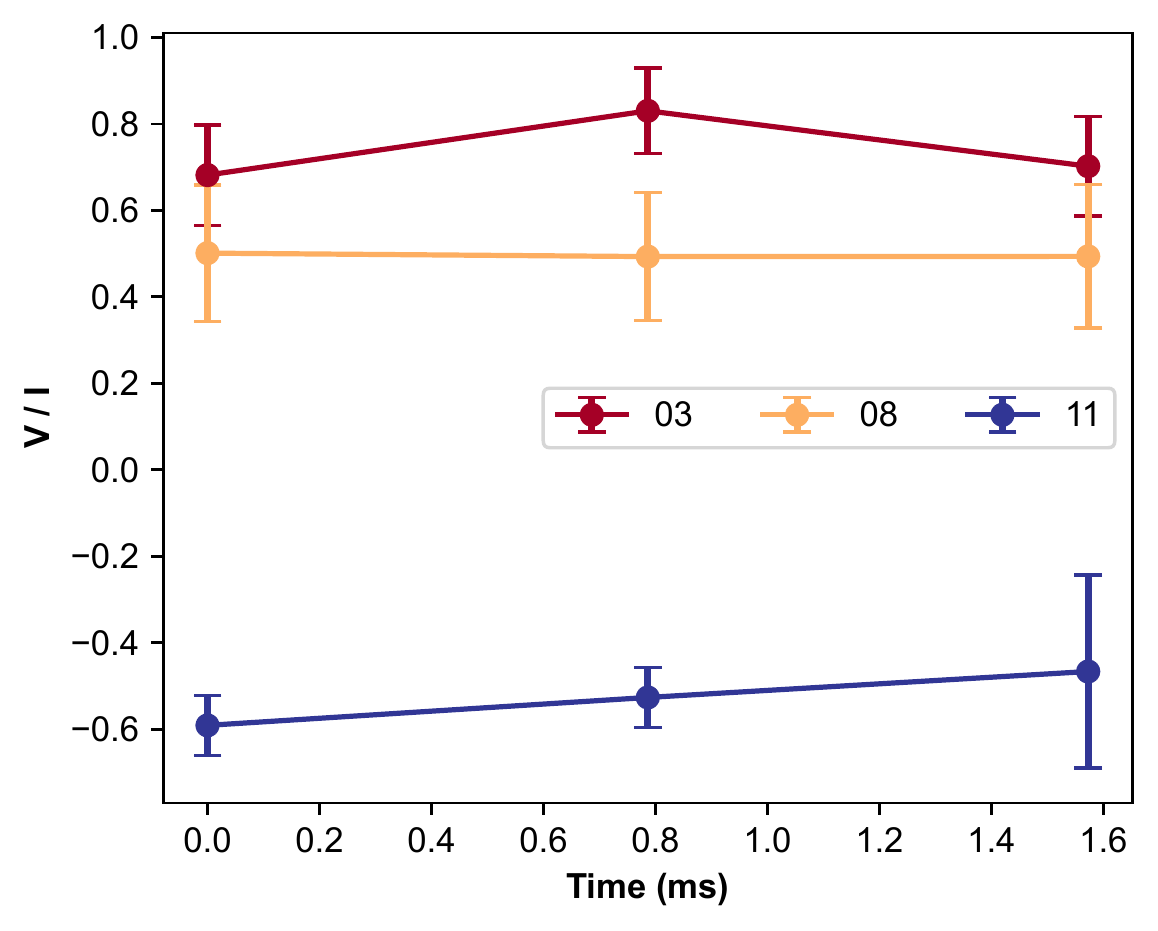}
  \caption{Degrees of circular polarization across $\sim$2\,ms of burst 3, 8, 11 of FRB~20121102A. The variations are within the ranges of error bars.}
  \label{fig:change}
\end{figure}
